\documentclass[reprint, amsmath,amssymb,aps, physrev,]{revtex4-2}
\usepackage{amsmath,amssymb,graphicx,xcolor}
\usepackage{epstopdf}

\begin{document}

\preprint{}

\title{\textbf{ Hall effect in viscous flows of two-dimensional electrons
 \\ in samples with edges of arbitrary roughness }}

\author{A.\,V.~Gert}
\email{anton.gert@mail.ioffe.ru}
\author{P.\,S.~Alekseev}
\email{pavel.alekseev@mail.ioffe.ru}
\affiliation{Ioffe Institute, St. Petersburg 194021, Russia}

\date{\today}

\begin{abstract}

In ultra-clean conductors, fast inter-particle collisions can lead to the formation of a viscous electron fluid and realization of the hydrodynamic transport regime. Here we develop a theory of hydrodynamic magnetotransport of two-dimensional (2D) electrons in samples with low densities of defects and edges of arbitrary roughness. Within our model the roughness is described by a single parameter with the dimension of speed in the boundary conditions on sample edges. The electron-fluid flow profiles in long samples, as well as the corresponding longitudinal and Hall resistances, are calculated. The contribution to the Hall resistance associated with the relaxation processes exhibits a saturation in the limit of high magnetic field and a minimum as a function of the magnetic field for sufficiently rough edges. The minimum disappears as the edge roughness decreases or  the sample width and bulk scattering by defects increase. These properties of the Hall resistance can serve as the signs to identify the hydrodynamic regime of electron transport in experiments and can be used to determine its parameters, in particular, the degree of edge roughness.

\end{abstract}

\maketitle

\section{Introduction} The hydrodynamic regime of electron transport can be formed in high-quality conductors in which the electron-electron (ee) collision rate exceeds the rates of electron scattering by impurities, making viscosity substantial. Such a regime has been experimentally discovered in ultra-clean samples of graphene, GaAs quantum wells, and some other materials~\cite{Moll,Bandurin2016,Alekseev2016,Polini2020,Gusev2018,Wang2022}. In graphene samples, hydrodynamic flow of two-dimensional (2D) electrons was identified for the first time by the negative absolute resistance, induced by formation of whirlpools, in complex-shaped samples between some contacts~\cite{Bandurin2016,Torre2015,Levitov2016,Polini2020}. In GaAs quantum wells, the hydrodynamic transport regime, linear in current, was detected~\cite{Alekseev2016,Gusev2018,Wang2022} by observing strong negative magnetoresistance~\cite{Bockhorn2011,Hatke2012,Mani2013,Shi2014,Gusev2018,Wang2022}, arising from the magnetic-field dependence of electron viscosity~\cite{Gurzhi_J68,Alekseev2016}. Later, the hydrodynamic regime in these systems was also identified by observing the effects of the complex sample geometry on its resistance~\cite{Keser2021}.

In the last ten years, a large number of experiments have been performed to study various properties of the hydrodynamic electron transport (see Refs.~\cite{Sulpizio2019,Ku2020,Bockhorn2024,Wang2023,EstradaAlvarez2025,Sarypov}). For example, in recent work~\cite{Sarypov} the flow of a 2D electron fluid through a point contact in GaAs quantum-well structures with negligible sticking of the flow to the contact edges was observed. The theory of hydrodynamic transport has also been developed in many directions. In particular, high-frequency effects~\cite{Pellegrino2017,Alekseev2018,Moessner2018,AlekseevAlekseeva2019PRL,Afanasiev2023,AlekseevAlekseeva2025}, spin effects~\cite{Glazov2022,Afanasiev2022Rotational,Denisov2022,Zohrabyan2026,Denisov2023,Raichev2025,Alekseev2026Zeeman}, and the crossover between ballistic and hydrodynamic regimes~\cite{Guo2017,Holder2019,Afanasiev2021} have been studied.

In hydrodynamic electron systems, the Hall effect is of particular interest. Experimental observations of a specific hydrodynamic contribution to the Hall resistance in the electron flow were reported for graphene samples and GaAs quantum wells~\cite{Berdyugin,Gusev2,Raichev2}. In Ref.~\cite{Berdyugin} the Hall viscosity of 2D electrons was determined from measurements of the Hall effect in asymmetric graphene samples. In Refs.~\cite{Gusev2,Raichev2} a mixed ballistic-hydrodynamic regime of the 2D electron transport in high-quality GaAs quantum wells was realized, and it was shown  that the measured Hall resistance  contains a substantial contribution from the Hall viscosity.

Theoretical studies~\cite{Alekseev2016,Scaffidi2017} have investigated the hydrodynamic contribution to the Hall effect in the flow of 2D electrons in long symmetric samples, caused by Hall viscosity. Such a contribution originates from the Hall field in the sample bulk and  is proportional to the squared ratio of the ee-scattering length  to the sample size. In addition to this bulk contribution, the near-edge layers can produce a contribution to the Hall effect of comparable magnitude~\cite{Afanasiev2022,Grigorev2025}. In these layers, the electron flow is semiballistic because it is determined by both ee-collisions and the scattering of 2D electrons at the sample edges. In Refs.~\cite{Holder2019,Raichev2,Afanasiev2021,Raichev2022,Raichev2} a unified theory of the ballistic-hydrodynamic magnetotransport at various boundary conditions was developed. It is based on a straightforward solution of the kinetic equation in both the sample bulk and the edge vicinities. In particular, the numerical solution of the kinetic equation in Ref.~\cite{Raichev2} explains the longitudinal magnetoresistance and the Hall resistance observed in~\cite{Gusev2,Raichev2}.

The type of sample edges plays a crucial role in determining corrections to the Hall effect, as edges with different degrees of roughness can induce adhesion, reflection, and/or slip of 2D electrons at the boundaries. The influence of edge layers on the effects of hydrodynamic and ballistic-hydrodynamic electron transport was theoretically studied, for example, in Refs.~\cite{Kiselev2019,Raichev2022,Raichev2,Raichev2023,Afanasiev2025}. Often in these studies the scattering of electrons at sample edges was considered to be diffuse, that is,  when 2D electrons are reflected from the rough edges in all directions with equal probabilities. In Refs.~\cite{Kiselev2019,Raichev2022,Raichev2023} it was shown that an approximate solution of the kinetic equation in semiballistic near-edge layers can be represented by refined boundary conditions for the Navier-Stokes equation containing the hydrodynamic velocity and the slip length.
The dependence of the slip length on the magnetic field was derived from the solution of the kinetic equation in the near-edge layer in Refs.~\cite{Raichev2023}.

In Refs.~\cite{Guo2017,Afanasiev2025}, a simplified, but more universal approach was developed to describe the semiballistic flows in the near-edge layers: a spatially inhomogeneous term, that sharply increases near the sample edges and models electron scattering of arbitrary strength at the boundaries, was introduced into the collision integral of the kinetic equation. This method describes not only specular and diffuse scattering of 2D electrons at sample edges, but also the entire range of boundary conditions, from complete sticking of the fluid to the edges to complete slip. It also allows us to determine the analytical dependence of the slip length on edge roughness, the magnetic field, and a flow frequency~\cite{Afanasiev2025}.

In this work, based on the approach of Refs.~\cite{Guo2017,Afanasiev2025}, we develop an analytical theory of stationary hydrodynamic magnetotransport of 2D electrons in samples with arbitrary edge types and weak residual defects in the bulk. We solve the Navier-Stokes equations for the hydrodynamic velocity and Hall electric field with the boundary conditions obtained in Ref.~\cite{Afanasiev2025}. First, we show that the giant negative magnetoresistance, which is realized in sufficiently narrow and clean samples with rough edges, gradually disappears as the edge smoothness increases. Second, we demonstrate that, at sufficiently rough edges, the relaxation contribution to the Hall resistance, $\delta \rho_{xy} =\rho_{xy} - \rho_{xy}^{st} $, changes non-monotonically with the magnetic field~$B$ and exhibits a saturation at high fields (here $\rho_{xy}^{st}  $ is the standard ``kinematic'' Hall resistivity related to the balance of the electric and magnetic Lorentz forces). The position of the minimum and the saturated value of this dependence~$\delta \rho_{xy}(B)$ are determined by the degree of edge roughness, momentum relaxation by the bulk defects, and the sample width. With increasing slip at the edges, the sample width, and the scattering in the bulk, the dependence~$\delta \rho_{xy}(B)$ becomes monotonic and its magnitude gradually decreases. We also obtain a universal relation between the magnetic-field dependencies of the longitudinal and Hall resistances within this model.

Experimental observation of all these properties of the Hall resistance can be used as an additional  method to identify the hydrodynamic regime of transport and characterize the ultra-pure samples where it is realized.

\section{Model} 
\begin{figure}
  \includegraphics[width=0.5\textwidth]{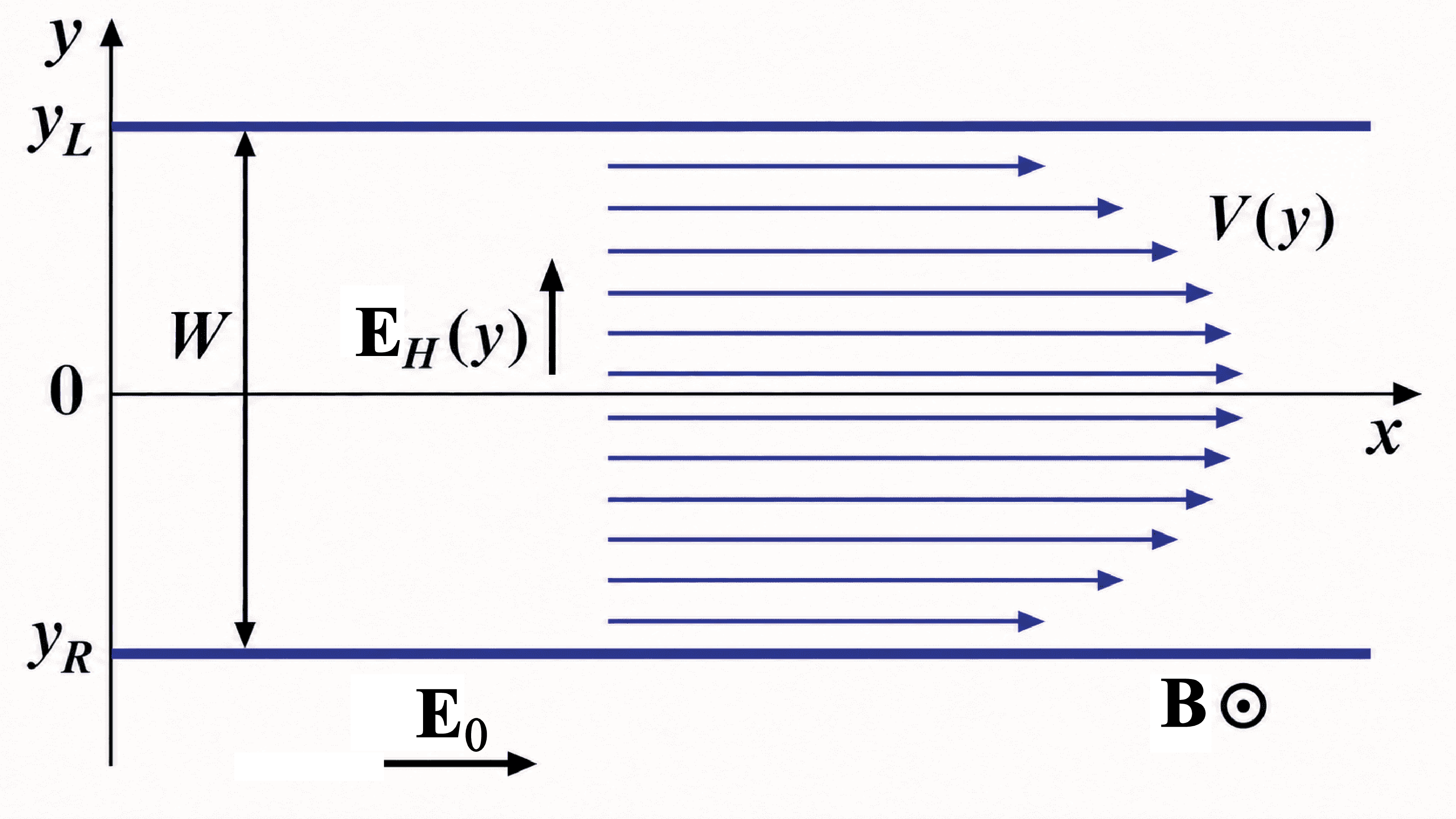}
  \caption{ Flow of a 2D electron fluid in a long sample with moderately rough edges, corresponding to substantial slip of the fluid flow at the edges. The following quantities are indicated: the external electric field $\mathbf{E}_0$; the magnetic field $\mathbf{B}$, perpendicular to the 2D electron layer; Hall electric field $ \mathbf{E}_H(y) $, arising to compensate for the magnetic Lorentz force in the $y$ direction; the velocity profile of the electron flow~$V_x (y) = V (y)$.}
  \label{scheme}
\end{figure}

We consider flows of 2D electrons in long samples bounded by the longitudinal edges at~$y_{L} = W/2$ and~$y_{R} = - W/2$, with the length~$L$ much larger than the width~$W$: $L \gg W$ (see Fig.~\ref{scheme}). The magnetic field $\mathbf{B} = B  \, \mathbf{e}_z$ is applied perpendicular to the 2D layer and an external electric field $E_0 \, \mathbf{e}_x$ is applied along the sample (by the bias between the contacts). The electron flow is directed along the $x$ axis: $\mathbf{V}(y) = V(y) \, \mathbf{e}_x$. Hall electric field $E_H(y) \, \mathbf{e}_y$ arises perpendicular to the flow due to a perturbation of the electron density~$\delta n(y)$, in order to compensate the magnetic Lorentz force~$(e/c)[ \mathbf{V}(y) \times \mathbf{B}]$.

Our aim is to study flows in such a system for arbitrary types of longitudinal edges: (i) nearly perfectly smooth edges, corresponding to nearly specular electron scattering; (ii) moderately rough edges, corresponding to the diffuse scattering (electrons reflected from the edge are scattered into all angles with equal probability); and (iii) extremely rough (or irregular) edges, corresponding to almost complete sticking of the electron fluid to the edges. In Ref.~\cite{Afanasiev2025} the theory of high-frequency electron flows in the purely hydrodynamic approximation in samples with arbitrary edges was developed based on the results of Refs.~\cite{Guo2017,Alekseev2019}. In the present work, we use the approach of~\cite{Afanasiev2025} to study stationary magnetotransport, especially, the Hall effect in this system.

The relaxation of the perturbed part of the electron distribution function~$\delta f$ in the bulk of the sample can be described by the simplified collision integral~\cite{Guo2017,Afanasiev2025}:
\begin{equation}
\label{int1}
  St_{\mathrm{b}}[\delta f] =
  - \gamma_{ee} \sum_{|m|\geq 2} \hat{P}_m[\delta f]
  - \gamma_{imp} \sum_{|m|\geq 1} \hat{P}_m[\delta f]
  \, ,
\end{equation}
where $\hat{P}_m$ are projectors onto the harmonics of the distribution function~$\delta f (y,\varphi)$ proportional to $e^{i m \varphi}$ (here $\varphi$ is the angle of the velocity of 2D electrons); $\gamma_{ee} = 1 / \tau_{ee}$ is the relaxation rate due to momentum conserving ee-collisions; $\gamma_{imp} = 1 / \tau_{imp}$ is the relaxation rate due to scattering by bulk defects, leading to momentum relaxation. We assume $\tau_{ee} \ll \tau_{imp}$, as it is typical for clean ``hydrodynamic'' samples. Operator~(\ref{int1}) leads to fast relaxation of the shear stresses~$\sigma_{ik}$ associated with the harmonics~$|m| = 2$ and to slow relaxation of the hydrodynamic velocity~${V}$, which is proportional to the harmonics~$|m|=1$.

The relaxation of the electron distribution at the edges in the long sample (see Fig.~1) is described by a local collision operator containing, in particular, the terms that lead to relaxation of the $|m|=1$ harmonics~\cite{Guo2017,Afanasiev2025}:
\begin{equation}
 \label{St_ed}
\begin{gathered}
  St_{\mathrm{ed}} ^{(1)} [\delta f]
  \: =\:
   - \: a \, [ \, \delta(y - y_L)
 \:+
\\
  + \: \delta(y - y_R) \, ] \: (\, \hat{P}_{m=1}[\delta f] \,
  + \, \hat{P}_{m=-1}[\delta f] \,)
  \: .
\end{gathered}
\end{equation}
Here the parameter $a$ characterizes the efficiency of electron scattering at the edges. As it was shown in~\cite{Afanasiev2025}, collision integral~(\ref{St_ed}) with intermediate values of $a$, $a \sim v_F$, approximately describes diffuse reflection of electrons from rough edges; the limit $a \gg v_F$ corresponds to very rough edges, leading to almost complete sticking of the electron flow, $V|_{y=y_L,y_R} = 0$; the limit $a \ll v_F$ corresponds to smooth edges with nearly specular electron scattering, which gives the boundary condition of a zero normal derivative of the velocity: $(dV/dy)|_{y=y_L,y_R} = 0$ (for proof of these two boundary conditions within our model see the text below).

\begin{figure*}
  \includegraphics[width=\textwidth]{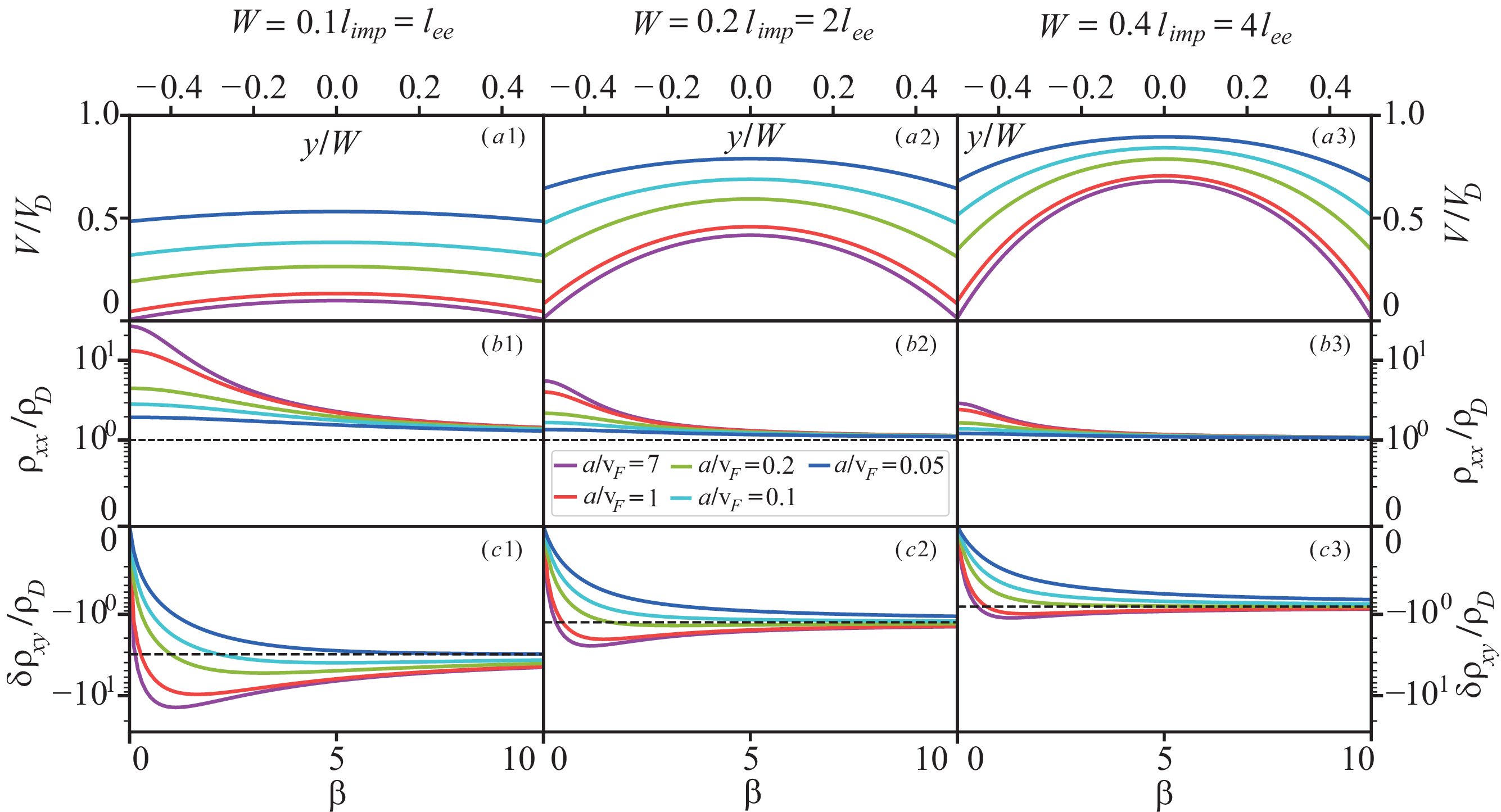}
  \caption{Characteristics of the flow of the 2D electron fluid in long samples with three widths~$W$ and various degrees of edge roughness, described by the parameter~$a$. The columns correspond to different ratios of the sample width to the relaxation lengths~$l_{imp} = v_F \tau_{imp}$ and $l_{ee} = v_F \tau_{ee}$, indicated at the top. The parameter $\xi_0=\lambda_0 W/2$, governing the relative contributions of the Ohmic and hydrodynamic components at zero magnetic field, takes the values $0.16$, $0.32$, and $0.63$ for the three columns. The top row [panels~(a1)--(a3)] shows the hydrodynamic velocity profiles~$V(y)$ on a linear scale (see Fig.~1). The middle row [panels~(b1)--(b3)] presents the mean sample longitudinal resistivities~$\rho_{xx} = E_0/(I/W)$ on a logarithmic scale in units of~$\rho_D = m/(ne^2\tau_{imp})$ versus the dimensionless magnetic field~$\beta = 2 \omega_c \tau_{ee}$. The bottom row [panels~(c1)--(c3)] shows the hydrodynamic ~$ \delta \rho_{xy} = \rho_{xy} - \rho_{xy} ^{st}$ contributions to the averaged Hall resistivities~$\rho_{xy} = U_H / I$ in units of~$\rho_D$ on a logarithmic scale versus~$\beta$. Each panel presents results for different types of sample edges (curves of different colors): nearly smooth edges, nearly complete slip~($a \ll v_F$); very rough edges, nearly complete sticking~($a \gg v_F$); and intermediate edge roughness corresponding to the diffuse electron reflection~($a \sim v_F$).}
  \label{pic}
\end{figure*}

In the simplified model of stationary flows developed here we assume, following Ref.~\cite{Afanasiev2025}, that the distribution function has a purely hydrodynamic form, that is, it contains only the first and second harmonics:
\begin{equation}
 \label{f_hydr}
 \delta f(y, \varphi)
\: =
  \sum_{m=\pm1, \pm2} \delta f_m(y) \, e^{i m \varphi}
\:.
\end{equation}
Criteria for the applicability of this approximation are discussed in Ref.~\cite{Afanasiev2025}\footnote{In real samples, contributions $\delta f_{\mathrm{non\mbox{-}hydr.}}$ in the distribution function from higher angular harmonics, $m\geq3$, which describe a semiballistic flow in the near-edge layers, can be important. Note that Ref.~\cite{Afanasiev2022} was examined the role of such contributions at weak magnetic fields for a Poiseuille-flow model with diffuse reflection from the edges by considering a simplified distribution function with harmonics~$|m|=3$. In the model of~\cite{Afanasiev2022}, the boundary conditions were obtained using the variational principle, proposed there: by minimizing the residual functional associated with satisfying the exact boundary conditions for general distribution functions, when the simplified distribution function with harmonics~$|m|=3$ is substituted into them. Studying the contributions $\delta f_{\mathrm{non\mbox{-}hydr.}}$ within the model of the present work, with a nonlocal edge-scattering operator of the type of Eq.~(\ref{St_ed}) with~$\hat{P}_m$, $|m| \geq 2 $, is also of interest and may lead to a significant refinement of the results presented here.}.

In the bulk of the sample, the kinetic equation with the collision integral~(\ref{int1}) for a distribution function of the form~(\ref{f_hydr}) leads to the Navier-Stokes equations for the hydrodynamic velocity~$V_x \equiv V(y,t)$ and Hall electric field $E_y \equiv E_H(y,t)$ (see, for example,~\cite{Alekseev2016}). For stationary flows under a dc driving field~$E_0$, the Navier-Stokes equations for the longitudinal velocity~$V(y)$ and Hall field~$E_H(y)$ in the presence of the magnetic field~$\mathbf{B} = B \, \mathbf{e}_z$ can be written as~\cite{Alekseev2016}:
\begin{equation}
\label{sys}
\left\{
\begin{gathered}
\displaystyle
  \frac{e E_0}{m} \, = \, - \, \eta_{xx} \, \frac{d^2 V}{dy^2}
  \,+ \, \frac{V}{\tau_{imp} }
\:,
\\
\displaystyle
 - \frac{e E_H}{m} \, = \, \omega_c V \,+\, \eta_{xy} \frac{d^2 V}{dy^2}
\:,
\end{gathered}
\right.
\end{equation}
where $m$ is the electron mass, $\omega_c = eB/(mc)$ is the electron cyclotron frequency, and~$\eta_{xx}$ and~$\eta_{xy}$ are the diagonal and off-diagonal components of the viscosity tensor of 2D electrons in the magnetic field:
\begin{equation}
 \label{eta_xx}
  \eta_{xx} \, = \, \frac{v_F^2\tau_{ee} / 4 }{1 + \beta^2}
 \: ,
 \quad
  \eta_{xy} \, = \, \beta \, \eta_{xx}
 \:.
\end{equation}
Here $v_F$ is the Fermi velocity; $\tau_{ee}$ is the shear stress relaxation time due to ee-collisions; $\beta = 2 \omega_c \tau_{ee}$ is the dimensionless magnetic field.  The quantity $\eta_{xy}$ is called the Hall viscosity and provides a hydrodynamic contribution to the Hall resistance in asymmetric samples~\cite{Berdyugin}, as well as in symmetric ones~\cite{Scaffidi2017,Afanasiev2022}.

Integrating the kinetic equation with collision integral~(\ref{St_ed}) over the edge regions $y \approx y_{L,R}$ gives the boundary conditions~\cite{Afanasiev2025}:
\begin{equation}
\label{bnds}
   \left(\mp l_{sl} \frac{dV}{dy} + V\right)
   \bigg|_{y = y_{L,R}} = \, 0
   \,,
\end{equation}
where $l_{sl} = l_{sl}(\omega_c)$ is the slip length, which depends on the magnetic field:
\begin{equation}
\label{l_sl}
   l_{sl} \, = \, \eta_{xx} (\omega_c) \, / \, a
  \:.
\end{equation}
It follows from Eqs.~(\ref{bnds}) and~(\ref{l_sl}) that the limit $a \to 0$ indeed corresponds to specular reflection and zero velocity gradient, $(dV/dy)|_{y=y_{L,R}} = 0$, whereas the limit $a \to \infty$ describes complete sticking of the fluid to the edges, $V|_{y=y_{L,R}} = 0$. In Ref.~\cite{Afanasiev2025} it was shown that the diffuse electron reflection from the edges corresponds to the estimate $a \sim v_F$, and in this case the slip length is of the order of the relaxation length $l_{ee}$ divided by a factor increasing with the magnetic field: $l_{sl} \sim l_{ee} / (1+\beta^2 )$.

In Refs.~\cite{Raichev2022,Raichev2023} the boundary conditions with the magnetic-field-dependent slip length, similar to result~(\ref{l_sl}), were obtained and compared with the result for~$l_{sl}$ from the numeric solution of the kinetic equation in the near-edge regions. It was assumed that arbitrary angle-dependent reflection probabilities for 2D electrons scattered at sample edges, which was expressed through an integral boundary condition for the distribution function. Our approach based on Eqs.~(\ref{St_ed}) is much simpler, however, it is less rigorous. As a result, it allows us to describe via a single parameter~$a$ any type of the sample edges, from specular to diffuse and very rough (irregular). The last case lies beyond the applicability of kinetic equation approaches used in Refs.~\cite{Raichev2022,Holder2019,Scaffidi2017,Afanasiev2021}. In our model, the diffusive, very rough and irregular, edges correspond to a large parameter $a \gg v_F$, whereas kinetic equation approaches for 2D electrons with angle-dependent scattering do not involve any similarly large velocity parameter.

\section{Results} The solution of the first equation in system~(\ref{sys}) with boundary conditions~(\ref{bnds}) is given by the formula:
\begin{equation}
\label{V}
		 V (y) \, = \, V_D \, \Big[ 1 -
  \frac{\cosh(\lambda y) / \cosh(\lambda W/2) }
  { 1 + \lambda l_{sl} \tanh ( \lambda W/ 2))} \Big]
  \:,	
\end{equation}
where $\lambda = \sqrt{1/(\eta_{xx}\tau_{imp})} = 1/l_G$ is the characteristic length scale over which the velocity amplitude $V(y)$ decays [the value $l_G = \sqrt{\eta_{xx}\tau_{imp}}$ is called the Gurzhi length], and $V_D= eE_0\tau_{imp}/m$ is the Drude drift velocity. The parameter~$\xi( B ) = \lambda W/ 2 $ controls the relation between the Ohmic and the hydrodynamic contributions in the whole flow.

Figures~\ref{pic}(a1)--(a3) show the obtained velocity profiles for samples of different widths and different degrees of edge roughness. As the ratio of the sample width~$W$ to the Gurzhi length~$l_G$ changes, the flow profile changes from parabolic to nearly flat. The sticking parameter $a$ also affects the velocity amplitude. The velocity is highest in the widest samples with the smallest~$a$.

The following parameters were used for our calculations: the 2D electron density $n = 9 \cdot 10^{11}$~cm$^{-2}$, corresponding to the Fermi velocity $v_F = 4 \cdot 10^{7}$~cm/s for GaAs and the interaction parameter $r_s=0.59$; the temperature $T = 30$~K, corresponding to the shear-stress relaxation time due to ee-collisions, $\tau_{ee}=2 \cdot 10^{-12}$~s, according to the theory of shear viscosity in a 2D electron Fermi gas presented in Ref.~\cite{Alekseev2020}; the relaxation time of scattering by the bulk defects $\tau_{imp}=10\tau_{ee}$. Such parameters are characteristic for the hydrodynamic transport regime in ultra-clean GaAs quantum wells studied in Refs.~\cite{Gusev2018,Gusev2}.

Note that, for these parameters, the left panels of Fig.~2 correspond to the case where the shear-stress relaxation length is equal to the sample width, $W = l_{ee}$, placing the system at the boundary of applicability of the hydrodynamic description, which requires the inequality $W \gg l_{ee}$. For such an intermediate regime, where $l_{ee}\sim W$, the hydrodynamic and ballistic contributions to all transport effects play comparable roles (see also Ref.~\cite{Afanasiev2021}, which studied both these two contributions to magnetotransport in the opposite limit, $W \ll l_{ee } $). The remaining, central and right, panels in Fig.~2 show results for wider samples with the same relaxation lengths, so that the inequality $W\gg l_{ee}$ is satisfied with some margin.

The total electric current through the structure, $I = e n \int_{y_R}^{y_L} V(y) \,dy$, takes the form:
\begin{equation}
\label{I0}
	 I \, = \, I_D \, \Big[ 1 -
  \frac{\tanh (\xi ) \, / \, \xi}
  { 1 + \lambda \, l_{sl} \tanh ( \xi )} \Big]
  \:,	
  \quad
   \xi =  \lambda W/2
  \:,
\end{equation}
where $I_D=en V_D W$ is the Drude value of the current. The averaged sample resistivity is defined as~$\rho_{xx} = E/(I/W)$ and is given by the formula:
\begin{equation}
\label{I}
  \rho_{xx} = \rho_D \, \frac{1 \, + \, s\,F(\xi) }{1 \, + \, (s-1)\,F(\xi) }
  \,, \qquad
  F(\xi ) = \frac{\tanh ( \xi ) }{ \xi}
\,,
\end{equation}
where $\rho_D = m/(ne^2\tau_{imp})$ is the Drude resistivity; $s=W/(2\tau_{imp} a)$, is the parameter characterizing the relative intensity of the momentum relaxation at the edges compared with the momentum relaxation in the bulk.

In Figures~\ref{pic}(b1)--(b3) we plot the dependence of the dimensionless longitudinal magnetoresistance $\rho_{xx} / \rho_D = I_D/I$ on the parameter $\beta = 2 \omega_c \tau_{ee}$, proportional to a magnetic field. For the narrow samples with rough edges, for which $\lambda |_{B=0} W \ll 1 $, the magnetoresistance decreases with the magnetic field because in this regime it is proportional to the viscosity coefficient $\eta_{xx}(\beta) \sim 1/(1+\beta^2)$~(\ref{eta_xx})~\cite{Alekseev2016}. In wide samples ($\lambda |_{B=0} W \gg 1 $) and/or at a large slip the electron transport is nearly Ohmic and the resistivity $\rho_{xx}$ depends weakly on the magnetic field. In this case, the resistivity is close to $\rho_D$, determined by electron momentum relaxation due to the scattering by bulk defects. As the magnetic field increases, the width of the near-edge layers $l_G(B)$ decreases monotonically, and the transport regime becomes nearly Ohmic also for narrow samples ($\lambda |_{B=0} W \ll 1 $). At a low magnetic field, the calculated resistance increases with increasing edge roughness at any~$W$ because of the additional momentum relaxation due to electron scattering at the edges.

The Hall voltage~$U_H = - \int_{y_R}^{y_L} E_H(y) dy $ between the longitudinal edges is obtained from the second equation in system~(\ref{sys}). It can be written as $U_H = U_H^{st} + \delta U_H$, where $U_H^{st} = IB/nec$ is the standard Hall voltage (corresponding to the balance of the magnetic and electric Lorentz forces in the $y$ direction), and $\delta U_H$ is the hydrodynamic correction:
\begin{equation}
\label{bas}
	\delta U_H
  \, = \,
  (m/e) \, \eta_{xy} \, [ \, V'|_{y=y_L} - V'|_{y=y_R} \, ]
\:.
\end{equation}

The corresponding cross-section-averaged Hall resistivity, $\rho_{xy} = (U_H^{st} + \delta U_H)/I$, can also be written as the sum $\rho_{xy} = \rho_{xy}^{st} + \delta \rho_{xy}$, where $\rho_{xy}^{st} = B/(nec) = [\beta  \tau_{ imp} / (2 \tau_{ee})]\, \rho_D$ is the standard Hall resistivity corresponding to $U_H^{st}$, and $\delta \rho_{xy}$ is the relaxation-induced correction:
\begin{equation}
 \label{rho_xy__fin}
  \delta \rho_{xy} \,
  = \,
  - \,
   2 \, \beta \, \rho_D \, \frac{ F(\xi) }{1 \, + \, (s-1)\,F(\xi) }
   \:.
\end{equation}

It is important that both the longitudinal~(\ref{I}) and the Hall averaged resistivities~(\ref{rho_xy__fin}) depend on the edge scattering parameter~$a$ only via the combination:
\begin{equation}
 \label{s}
 s \,= \, W \, / \, (2\tau_{imp} a)
\:,
\end{equation}
describing the ratio of the rates of the momentum relaxation in the bulk, $1/\tau_{imp}$, and of the effective momentum relaxation at the edges,~$W/(2a)$.

In Ref.~\cite{Raichev2} a similar result for $\delta \rho_{xy}$ was obtained for a fixed slip length, $l_{sl}$. However, the full magnetic-field dependence of the Hall resistance $\delta \rho_{xy}(B)$ arising from the magnetic-field dependence of the slip length~(\ref{l_sl}), and the regime of very rough sample edges, corresponding to $a \gg v_F$, have not been studied previously.  In particular, the maximum of $\delta \rho_{xy}(B)$ at large edge roughness, $a>a_0$; its disappearance for $a<a_0$; the saturation of $\delta \rho_{xy}(B)$ at $B\to\infty$; and the edge-roughness dependence of $ \rho_{xx}$ and $\delta \rho_{xy}$ governed by the magnetic-field-independent parameter $s$~(\ref{s}) have not been reported previously.

\begin{figure}
  \includegraphics[width=0.4\textwidth]{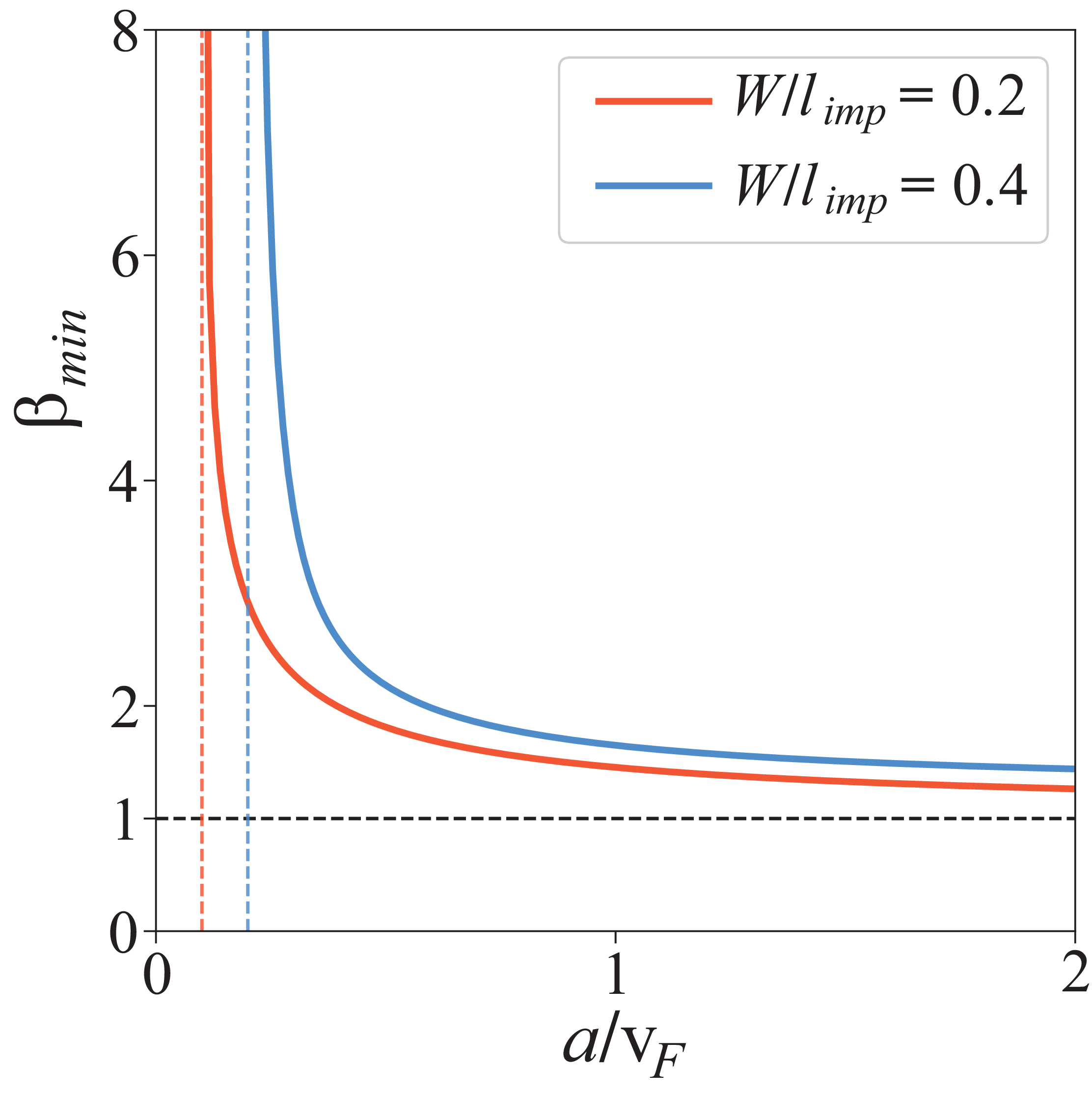}
  \caption{Position of the minimum~$\beta_{min}$ of the correction to the Hall resistance~$\delta \rho_{xy}(\beta)$ [see Fig.~2(c1)--(c3)] as a function of the parameter $a$ for samples of two different widths. The curves~$\beta_{min} (a) $ have vertical asymptotes at $a=a_0=W/(2\tau_{imp})$, corresponding to $s=[W/(2\tau_{imp})]/a=1$, at which the dependence $\delta\rho_{xy}(\beta)$ changes from monotonic to non-monotonic, with a minimum. As the parameter~$a$ increases, all curves saturate at~$\beta_{min}>1$, corresponding to the purely bulk result [Eq.~(\ref{rho_xy__fin}) at $s=0$]. }
  \label{min}
\end{figure}

In Figures~\ref{pic}(c1)--(c3) we present the dependence of the correction to the Hall resistance $\delta\rho_{xy}$ on the magnetic field. For smooth edges ($a\ll v_F$, more precisely $s>1$), the function $\delta \rho_{xy}(\beta)$ decreases monotonically to some finite value with increasing magnetic field. For rough edges, when $s<1$, the function $\delta \rho_{xy}(\beta)$ becomes non-monotonic and has a minimum at $\beta=\beta_{min}$, whose position depends on the sample width, edge roughness and relaxation lengths. For complete sticking, $a \gg v_F $, and the position of this minimum, $\beta_{min}^{\infty } >1$, is the value depending on the ratios $W/l_{ee}$ and $W/l_{imp}$.
 At weak scattering by bulk defects, when $W\ll l_G$ (and~$a \gg v_F $), equation~(\ref{rho_xy__fin}) yields $\delta \rho_{xy}(\beta)\propto\eta_{xy}(\beta)$, and the minimum occurs at~$\beta_{min}^{\infty}=1$.

Analysis of Eq.~(\ref{rho_xy__fin}) shows that, as~$a$ decreases, the minimum shifts toward larger magnetic fields and disappears at $a_0=W/(2\tau_{imp})$ [see Figures~\ref{pic}(c1)--(c3)]. Such value~$a_0$ corresponds to the equality $s=[W/(2\tau_{imp})]/a=1$. The dependence of the minimum position on the roughness parameter $a$ for different sample widths is shown in Fig.~\ref{min}. In accordance with the above, the curves $\beta_{min}(a)$ have a vertical asymptote as $a\to a_0$ and decrease monotonically with increasing $a$, approaching the limiting values $\beta_{min}^{\infty}>1$.

It is important that the above resulting equations for the longitudinal~(\ref{I})  and Hall~(\ref{rho_xy__fin}) resistances are closely related, namely one can be expressed in terms of the other by the relation:
\begin{equation}
  \label{connection}
    \rho_{xx} (\beta) \, + \, \frac{ \delta \rho_{xy} (\beta) }{2 \, \beta }
    \, \equiv \,  \rho_D
 \:.
\end{equation}
We can conclude from Eq.~(\ref{connection}) that the applicability of the current model, based on Eqs.~(\ref{St_ed}) and~(\ref{f_hydr}), can be checked by comparing the experimental data for the dependencies~$\rho_{xx} (B) $ and~$\rho_{xy} (B) $ with this equation.

\section{Discussion} Let us consider in detail the behavior of the obtained correction to the Hall resistance~$\delta \rho_{xy}$ in different limiting cases.

First, we analyze the flow regime when the viscosity plays a major role (the ``strongly hydrodynamic'' regime), $W \ll l_G$ and $\xi \ll 1$. For this regime and for very rough edges, $a\gg v_F$, $s\ll 1$, equations~(\ref{bas}) and~(\ref{I}) yield~\cite{Scaffidi2017,Afanasiev2022}:
\begin{equation}
 \label{abs}
  \delta \rho_{xy} ^{hydr}
  \, = \,
  - \, 12 \, \frac{\eta_{xy}}{\omega_c W^2}\,\rho_{xy}^{st}
 \:.
\end{equation}
It is seen that this result corresponds to the general formula~(\ref{rho_xy__fin}) in the limit~$s \ll 1$ and~$\xi \ll 1 $. In the low-field limit, $\beta \ll 1$, the purely hydrodynamic correction~(\ref{abs}) takes the form~\cite{Scaffidi2017,Afanasiev2022}:
\begin{equation}
 \label{Bto0}
  \delta \rho_{xy}^{hydr}
  \, = \,
  -\,6\,\frac{l_{ee}^2}{W^2}\,\rho_{xy}^{st}
\:.
\end{equation}
Equation~(\ref{Bto0}) directly shows that the correction $\delta\rho_{xy}$ is smaller than the standard value $\rho_{xy}^{st}$ by a factor of order $(l_{ee}/W)^2$, which is the square of the small parameter $l_{ee}/W\ll1$. This parameter governs the applicability of the hydrodynamic approximation. In the high-field limit, $\beta \gg 1 $ (also for infinitesimal parameters $s\to0$, $\xi\to0$), the purely hydrodynamic correction~(\ref{abs}) tends to zero as:

\begin{equation}
  \delta\rho_{xy} ^{hydr}
  \, = \,
  - \, \frac{3}{2} \, \frac{R_c^2}{W^2} \, \rho_{xy}^{st}
  \:,
\end{equation}
where $R_c=v_F/\omega_c$ is the cyclotron radius.

Second, let us consider the intermediate hydrodynamic-Ohmic regime, in which $ W\lesssim l_G$ and the diffuse reflection takes place ($a\sim v_F $ and herewith $s\lesssim1$), at the moderate magnetic fields, $\beta\lesssim1$. This case corresponds to the slip lengths of the order of the ee-scattering length: $l_{sl}(\beta)\sim l_{ee}$. In this regime the hydrodynamic correction~(\ref{rho_xy__fin}) remains of the same order of magnitude as the value in Eq.~(\ref{Bto0}):
\begin{equation}
\delta\rho_{xy} ^{int} \, \sim \, - \, \frac{ l_{ee}^2}{W^2} \, \rho_{xy}^{st}
\:.
\end{equation}
This result is consistent with Ref.~\cite{Afanasiev2022}, where the Hall resistance at low magnetic fields, $\beta \ll 1 $, and at the diffuse reflection from the edges was calculated in the two independent ways: by using variational boundary conditions for the correction to the hydrodynamic distribution function, $\delta f_{non-hydr.} \sim e^{\pm i3\varphi}$, and by an approximate solution of the full kinetic equation in the near-edge regions, $W/2-|y|\leq\sim l_{ee}$. We remind that the diffuse reflection of 2D electrons from moderately rough edges corresponds to the estimate: $a\sim v_F$~within the current model~\cite{Afanasiev2025}.

Third, let us consider in detail the limit of very large magnetic fields, $\beta\gg1$, at any fixed finite sample width~$W$, bulk defect relaxation time~$\tau_{imp}$, and edge roughness~$a$. In this case we always have $\xi \gg 1 $, since $\xi \sim \sqrt{1+\beta ^2 }$. The correction to the Hall resistance~(\ref{rho_xy__fin}) saturates with increasing magnetic field at the value:
\begin{equation}
 \label{rho_xy__fin_inf}
  \delta \rho_{xy} ^{\infty} \, = \,
 - \, \frac{ \sqrt{l_{ee} } \, R_c }{ 2 \, \sqrt{l_{imp}} \: W} \, \rho_{xy} ^{st}
    =
-  \frac{ \, \sqrt{l_{imp} \, l _{ee} \, }}{ 2 \,W } \, \rho_D
 \:,
\end{equation}
which does not contain the parameter~$a$, that is, it is independent of the edge type.  This quantity is small compared with $\rho_{xy}^{st}$ in the hydrodynamic regime characterized by a short ee-scattering length $l_{ee}$ and a relatively large impurity scattering length $l_{imp}$, for example, when $l_{ee}\ll W$ and $l_{ee}\lesssim l_{imp}$. The values~$\delta \rho_{xy} ^{\infty} $~(\ref{rho_xy__fin_inf}) at different~$W,\,l_{ee,\,imp}$ are the horizontal asymptotes in plots of~$\delta \rho_{xy} (\beta)$ in Fig.\ref{pic}~(c1-c3).

The absence of the minimum \(\beta_{\min}\) in the dependence $\rho_{xy}(\beta)$ at the smooth edges, when~$s(a)>1$, and its appearance for rough edges corresponding to $s(a)<1$, as well as the monotonic behavior of $\beta_{min}(a)$, can be used to characterize samples. Namely, from the position \(\beta_{\min}(a)\)  of the minimum and from the independently known relaxation times $\tau_{imp}$ and $\tau_{ee}$ for a given sample, the parameter $a$, describing the edge roughness, can be determined. Measurements of the limiting values~(\ref{rho_xy__fin_inf}) of the dependence~$\delta \rho_{xy} (\beta) $ can provide an additional method to determine the ratios~$l_{ee}/W$ and~$l_{imp} / W$.

Next, we discuss the available experimental and theoretical results on the Hall effect in high-quality GaAs quantum wells.

A ballistic-hydrodynamic flow regime of a 2D electron fluid in  magnetic field was examined experimentally in high-purity GaAs quantum wells in ~\cite{Gusev2}; similar experimental results were presented in Ref.~\cite{Raichev2}. Along with the longitudinal magnetoresistance, the Hall resistance was measured in~\cite{Gusev2}. Apparently, the hydrodynamic and ballistic effects made comparable contributions to the  electron transport  in moderate magnetic fields~$\beta \lesssim 1 $, since the relation~$l_{ee}\sim W$ was established in~\cite{Gusev2}.

It is known that in very pure and narrow samples, for which~$W \ll l_{ee},\,l_{imp} $, the hydrodynamic contribution in transport becomes important only at sufficiently high magnetic fields, when the diameter of the electron cyclotron trajectory~$2R_c$ is smaller than the sample width~$W$~\cite{Scaffidi2017,Holder2019,Afanasiev2021}. At low magnetic fields, when~$R_c \gg W$, the ballistic effects due to the electron scattering at the edges dominate in magnetotransport.

In experiment~\cite{Gusev2} a positive correction~$\delta\rho_{xy}=\rho_{xy}-\rho_{xy}^{st}>0$, to the standard Hall resistance $\rho_{xy}^{st}=B/(nec)$ was observed at low fields, when $R_c \gtrsim W $. Apparently, this correction was induced mainly by the ballistic effects due to electron scattering at the edges, similarly as it takes place for the very pure and narrow samples,~$W \ll l_{ee,\,imp} $,  within theories~\cite{Scaffidi2017,Afanasiev2021}.
At higher magnetic fields, when $R_c \sim W$, the Hall resistance measured in~\cite{Gusev2} exhibited a negative correction to its standard value, $\delta\rho_{xy}=\rho_{xy}-\rho_{xy}^{st}<0$, that depended non-monotonically on the magnetic field. It reached a minimum at a certain magnetic field \(B_{\min}\), where \(R_c \sim l_{ee} \sim W\), and then rapidly increased toward zero as the magnetic field was increased further. In Ref.~\cite{Afanasiev2021}, such sign-alternating contribution~$\delta\rho_{xy}(B)$ was qualitatively explained within a ``quasi-ballistic'' model  in which ee-collisions were treated perturbatively, with predominant scattering at the sample edges. In Refs.~\cite{Scaffidi2017,Raichev2} a theory based on the numerical solution of the kinetic equation for 2D electrons in a long sample in the magnetic field at any relations between the parameters $W, \, l_{imp} , \, l_{ee}, \, R_c $ was developed and a rather well coincidence between the calculated Hall resistance~$\delta\rho_{xy}$ and the experimental results was obtained.

Since for the samples studied in experiments~\cite{Gusev2,Raichev2} both ballistic and hydrodynamic contributions to magnetotransport were apparently comparable, the negative  correction~$\delta\rho_{xy}(B)$ to the standard Hall resistivity~$\rho_{xy}^{st}$ observed at~$ \beta \sim 1 $  should contain a significant relaxation-induced contribution arising from the Hall viscosity.  Indeed, the hydrodynamic-Ohmic contribution  to the Hall resistivity~$\delta\rho_{xy} (B;a)$~(\ref{rho_xy__fin}), obtained above, and the purely hydrodynamic contribution~$\delta\rho_{xy} ^{hydr} (B)$~(\ref{abs}), obtained in Refs.~\cite{Scaffidi2017,Afanasiev2022}, are negative and have minima at magnetic fields corresponding to~$\beta \sim 1 $ or~$\beta \gg 1 $ (see Fig.~2). Next, the magnitude of the ratio~$\delta\rho_{xy} /\rho_{xy}^{st} $ in the range of parameters~$ W \gg l_{G} $, $s \sim 1$, and~$\beta \sim 1 $ is   estimated as~$l_{ee} ^ {3/2} / (W l_{imp}^{1/2})$. The standard temperature dependence of the ee-scattering length is~$l_{ee} \sim 1/T^2$~\cite{Alekseev2020}, so $\delta\rho_{xy} /\rho_{xy}^{st} $  rapidly decreases with temperature. The experimental data for this ratio presented in Ref.~\cite{Raichev2} also showed a rapid decrease with temperature. Thus, our result~(\ref{rho_xy__fin}) for the resistivity~$\delta\rho_{xy}  (B;a)$ qualitatively agrees with the experimental results of Refs.~\cite{Gusev2,Raichev2}.

Our analytic model can be expanded by accounting for several higher harmonics in the distribution function~$\delta f_{hydr.} + \delta f_{non-hydr.} $ in order to analytically describe also the ballistic contributions in the flows and the resulting values $\rho_{xx,xy}$ for any types of the samples and their edges, by a method similar to the one which was proposed in Ref.~\cite{Afanasiev2022} for the case of the diffuse edge scattering and low magnetic fields.
\\

\section{Conclusion} The hydrodynamic model of magnetotransport of a viscous 2D electron fluid in a long sample with residual defects and arbitrary types of edges has been developed. It provides a unified description of the viscous-Ohmic electron flows at realistic boundary conditions, ranging from slip to sticking of the flow at the sample edges. In particular, it has been shown that, as the sample width increases and the edges become smoother, the magnetic field dependence of the hydrodynamic contribution to the Hall resistance changes from a non-monotonic with a minimum to a monotonic one.  The position and magnitude of the minimum of this quantity can serve as an additional signature of the hydrodynamic transport regime and can provide a method to measure the parameters of the sample and its edges.

We thank G.~M.~Gusev for valuable discussions and comments.

This work was carried out under the state assignment of the Ministry of Science and Higher Education of the Russian Federation.

\bibliography{paper}

@article{Moll,
  author  = {Moll, P. J. W. and Kushwaha, P. and Nandi, N. and Schmidt, B. and Mackenzie, A. P.},
  title   = {Evidence for hydrodynamic electron flow in {PdCoO$_2$}},
  journal = {Science},
  volume  = {351},
  pages   = {1061--1064},
  year    = {2016},
  doi     = {10.1126/science.aac8385}
}

@article{Bandurin2016,
  author  = {Bandurin, D. A. and Torre, I. and Krishna Kumar, R. and Ben Shalom, M. and Tomadin, A. and Principi, A. and Auton, G. H. and Khestanova, E. and Novoselov, K. S. and Grigorieva, I. V. and Ponomarenko, L. A. and Geim, A. K. and Polini, M.},
  title   = {Negative local resistance caused by viscous electron backflow in graphene},
  journal = {Science},
  volume  = {351},
  pages   = {1055--1058},
  year    = {2016},
  doi     = {10.1126/science.aad0201}
}

@article{Polini2020,
  author  = {Polini, M. and Geim, A. K.},
  title   = {Viscous electron fluids},
  journal = {Phys. Today},
  volume  = {73},
  pages   = {28--34},
  year    = {2020},
  doi     = {10.1063/PT.3.4497}
}

@article{Levitov2016,
  author  = {Levitov, L. and Falkovich, G.},
  title   = {Electron viscosity, current vortices and negative nonlocal resistance in graphene},
  journal = {Nat. Phys.},
  volume  = {12},
  pages   = {672--676},
  year    = {2016},
  doi     = {10.1038/nphys3667}
}

@article{Alekseev2016,
  author  = {Alekseev, P. S.},
  title   = {Negative magnetoresistance in viscous flow of two-dimensional electrons},
  journal = {Phys. Rev. Lett.},
  volume  = {117},
  pages   = {166601},
  year    = {2016},
  doi     = {10.1103/PhysRevLett.117.166601}
}

@article{Gusev2018,
  author  = {Gusev, G. M. and Levin, A. D. and Levinson, E. V. and Bakarov, A. K.},
  title   = {Viscous electron flow in mesoscopic two-dimensional electron gas},
  journal = {AIP Adv.},
  volume  = {8},
  pages   = {025318},
  year    = {2018},
  doi     = {10.1063/1.5020763}
}

@article{Wang2022,
  author  = {Wang, X. and Jia, P. and Du, R.-R. and Pfeiffer, L. N. and Baldwin, K. W. and West, K. W.},
  title   = {Hydrodynamic charge transport in an {GaAs/AlGaAs} ultrahigh-mobility two-dimensional electron gas},
  journal = {Phys. Rev. B},
  volume  = {106},
  pages   = {L241302},
  year    = {2022},
  doi     = {10.1103/PhysRevB.106.L241302}
}

@article{Hatke2012,
  author  = {Hatke, A. T. and Zudov, M. A. and Reno, J. L. and Pfeiffer, L. N. and West, K. W.},
  title   = {Giant negative magnetoresistance in high-mobility two-dimensional electron systems},
  journal = {Phys. Rev. B},
  volume  = {85},
  pages   = {081304(R)},
  year    = {2012},
  doi     = {10.1103/PhysRevB.85.081304}
}

@article{Mani2013,
  author  = {Mani, R. G. and Kriisa, A. and Wegscheider, W.},
  title   = {Size-dependent giant-magnetoresistance in millimeter scale {GaAs/AlGaAs} 2D electron devices},
  journal = {Sci. Rep.},
  volume  = {3},
  pages   = {2747},
  year    = {2013},
  doi     = {10.1038/srep02747}
}

@article{Bockhorn2011,
  author  = {Bockhorn, L. and Barthold, P. and Schuh, D. and Wegscheider, W. and Haug, R. J.},
  title   = {Magnetoresistance in a high-mobility two-dimensional electron gas},
  journal = {Phys. Rev. B},
  volume  = {83},
  pages   = {113301},
  year    = {2011},
  doi     = {10.1103/PhysRevB.83.113301}
}

@article{Shi2014,
  author  = {Shi, Q. and Martin, P. D. and Ebner, Q. A. and Zudov, M. A. and Pfeiffer, L. N. and West, K. W.},
  title   = {Colossal negative magnetoresistance in a two-dimensional electron gas},
  journal = {Phys. Rev. B},
  volume  = {89},
  pages   = {201301(R)},
  year    = {2014},
  doi     = {10.1103/PhysRevB.89.201301}
}

@article{Keser2021,
  author  = {Keser, A. C. and Wang, D. Q. and Klochan, O. and Ho, D. Y. H. and Tkachenko, O. A. and Tkachenko, V. A. and Culcer, D. and Adam, S. and Farrer, I. and Ritchie, D. A. and Sushkov, O. P. and Hamilton, A. R.},
  title   = {Geometric control of universal hydrodynamic flow in a two-dimensional electron fluid},
  journal = {Phys. Rev. X},
  volume  = {11},
  pages   = {031030},
  year    = {2021},
  doi     = {10.1103/PhysRevX.11.031030}
}

@article{Bockhorn2024,
  author  = {Bockhorn, L. and Schuh, D. and Reichl, C. and Wegscheider, W. and Haug, R. J.},
  title   = {Influence of the electron density on the giant negative magnetoresistance in two-dimensional electron gases},
  journal = {Phys. Rev. B},
  volume  = {109},
  pages   = {205416},
  year    = {2024},
  doi     = {10.1103/PhysRevB.109.205416}
}

@article{Wang2023,
  author  = {Wang, Z. T. and Hilke, M. and Fong, N. and Austing, D. G. and Studenikin, S. A. and West, K. W. and Pfeiffer, L. N.},
  title   = {Nonlinear transport phenomena and current-induced hydrodynamics in ultrahigh-mobility two-dimensional electron gas},
  journal = {Phys. Rev. B},
  volume  = {107},
  pages   = {195406},
  year    = {2023},
  doi     = {10.1103/PhysRevB.107.195406}
}

@article{EstradaAlvarez2025,
  author  = {Estrada-{\'A}lvarez, J. and Salvador-S{\'a}nchez, J. and P{\'e}rez-Rodr{\'i}guez, A. and S{\'a}nchez-S{\'a}nchez, C. and Cleric{\`o}, V. and Vaquero, D. and Watanabe, K. and Taniguchi, T. and Diez, E. and Dom{\'i}nguez-Adame, F. and Amado, M. and D{\'i}az, E.},
  title   = {Superballistic conduction in hydrodynamic antidot graphene superlattices},
  journal = {Phys. Rev. X},
  volume  = {15},
  pages   = {011039},
  year    = {2025},
  doi     = {10.1103/PhysRevX.15.011039}
}

@article{Ku2020,
  author  = {Ku, M. J. H. and others},
  title   = {Imaging viscous flow of the {Dirac} fluid in graphene},
  journal = {Nature},
  volume  = {583},
  pages   = {537--541},
  year    = {2020},
  doi     = {10.1038/s41586-020-2507-2}
}

@article{Sulpizio2019,
  author  = {Sulpizio, J. A. and others},
  title   = {Visualizing {Poiseuille} flow of hydrodynamic electrons},
  journal = {Nature},
  volume  = {576},
  pages   = {75--79},
  year    = {2019},
  doi     = {10.1038/s41586-019-1788-9}
}

@article{Sarypov,
  author  = {Sarypov, D. I. and Pokhabov, D. A. and Pogosov, A. G. and Zhdanov, E. Yu. and Shevyrin, A. A. and Bakarov, A. K. and Shklyaev, A. A.},
  title   = {Slip electron flow in {GaAs} microscale constrictions},
  journal = {Phys. Rev. Lett.},
  volume  = {135},
  pages   = {236301},
  year    = {2025},
  doi     = {10.1103/PhysRevLett.135.236301}
}

@article{Pellegrino2017,
  author  = {Pellegrino, F. M. D. and Torre, I. and Polini, M.},
  title   = {Nonlocal transport and the {Hall} viscosity of two-dimensional hydrodynamic electron liquids},
  journal = {Phys. Rev. B},
  volume  = {96},
  pages   = {195401},
  year    = {2017},
  doi     = {10.1103/PhysRevB.96.195401}
}

@article{Alekseev2018,
  author  = {Alekseev, P. S.},
  title   = {Magnetic resonance in a high-frequency flow of a two-dimensional viscous electron fluid},
  journal = {Phys. Rev. B},
  volume  = {98},
  pages   = {165440},
  year    = {2018},
  doi     = {10.1103/PhysRevB.98.165440}
}

@article{Moessner2018,
  author  = {Moessner, R. and Sur{\'o}wka, P. and Witkowski, P.},
  title   = {Pulsating flow and boundary layers in viscous electronic hydrodynamics},
  journal = {Phys. Rev. B},
  volume  = {97},
  pages   = {161112(R)},
  year    = {2018},
  doi     = {10.1103/PhysRevB.97.161112}
}

@article{AlekseevAlekseeva2019PRL,
  author  = {Alekseev, P. S. and Alekseeva, A. P.},
  title   = {Transverse magnetosonic waves and viscoelastic resonance in a two-dimensional highly viscous electron fluid},
  journal = {Phys. Rev. Lett.},
  volume  = {123},
  pages   = {236801},
  year    = {2019},
  doi     = {10.1103/PhysRevLett.123.236801}
}

@article{Afanasiev2023,
  author  = {Afanasiev, A. N. and Alekseev, P. S. and Greshnov, A. A. and Semina, M. A.},
  title   = {Shear {Bernstein} modes in a two-dimensional electron liquid},
  journal = {Phys. Rev. B},
  volume  = {108},
  pages   = {235124},
  year    = {2023},
  doi     = {10.1103/PhysRevB.108.235124}
}

@article{AlekseevAlekseeva2025,
  author  = {Alekseev, P. S. and Alekseeva, A. P.},
  title   = {Highly correlated two-dimensional viscous electron fluid in moderate magnetic fields},
  journal = {Phys. Rev. B},
  volume  = {111},
  pages   = {235202},
  year    = {2025},
  doi     = {10.1103/PhysRevB.111.235202}
}

@article{Glazov2022,
  author  = {Glazov, M. M.},
  title   = {Valley and spin accumulation in ballistic and hydrodynamic channels},
  journal = {2D Mater.},
  volume  = {9},
  pages   = {015027},
  year    = {2022},
  doi     = {10.1088/2053-1583/ac3e04}
}

@article{Afanasiev2022Rotational,
  author  = {Afanasiev, A. N. and Alekseev, P. S. and Danilenko, A. A. and Greshnov, A. A. and Semina, M. A.},
  title   = {Rotational viscosity in spin resonance of hydrodynamic electrons},
  journal = {Phys. Rev. B},
  volume  = {106},
  pages   = {L041407},
  year    = {2022},
  doi     = {10.1103/PhysRevB.106.L041407}
}

@article{Denisov2022,
  author  = {Denisov, K. S. and Baryshnikov, K. A. and Alekseev, P. S.},
  title   = {Spin imaging of {Poiseuille} flow of a viscous electronic fluid},
  journal = {Phys. Rev. B},
  volume  = {106},
  pages   = {L081113},
  year    = {2022},
  doi     = {10.1103/PhysRevB.106.L081113}
}

@article{Zohrabyan2026,
  author  = {Zohrabyan, D. S. and Glazov, M. M.},
  title   = {Odd viscosity and anomalous {Hall} effect in two-dimensional systems with smooth disorder},
  journal = {JETP Lett.},
  volume  = {123},
  pages   = {257--264},
  year    = {2026},
  doi     = {10.1134/S0021364025610279}
}

@article{Denisov2023,
  author  = {Denisov, K. S. and Baryshnikov, K. A. and Alekseev, P. S.},
  title   = {Memory effects in the magnetoresistance of two-component electron systems},
  journal = {JETP Lett.},
  volume  = {118},
  pages   = {123--129},
  year    = {2023},
  doi     = {10.1134/S0021364023601860}
}

@misc{Alekseev2026Zeeman,
  author        = {Alekseev, Yu. O. and Alekseev, P. S. and Dmitriev, A. P.},
  title         = {Kinetic coefficients of two-dimensional electrons with strong {Zeeman} splitting},
  year          = {2026},
  eprint        = {2603.03105},
  archivePrefix = {arXiv},
  doi           = {10.48550/arXiv.2603.03105}
}

@article{Holder2019,
  author  = {Holder, T. and Queiroz, R. and Scaffidi, T. and Silberstein, N. and Rozen, A. and Sulpizio, J. A. and Ella, L. and Ilani, S. and Stern, A.},
  title   = {Ballistic and hydrodynamic magnetotransport in narrow channels},
  journal = {Phys. Rev. B},
  volume  = {100},
  pages   = {245305},
  year    = {2019},
  doi     = {10.1103/PhysRevB.100.245305}
}

@article{Afanasiev2021,
  author  = {Afanasiev, A. N. and Alekseev, P. S. and Greshnov, A. A. and Semina, M. A.},
  title   = {Ballistic-hydrodynamic phase transition in flow of two-dimensional electrons},
  journal = {Phys. Rev. B},
  volume  = {104},
  pages   = {195415},
  year    = {2021},
  doi     = {10.1103/PhysRevB.104.195415}
}

@article{Berdyugin,
  author  = {Berdyugin, A. I. and Xu, S. G. and Pellegrino, F. M. D. and Krishna Kumar, R. and Principi, A. and Torre, I. and Ben Shalom, M. and Taniguchi, T. and Watanabe, K. and Grigorieva, I. V. and Polini, M. and Geim, A. K. and Bandurin, D. A.},
  title   = {Measuring {Hall} viscosity of graphene's electron fluid},
  journal = {Science},
  volume  = {364},
  pages   = {162--165},
  year    = {2019},
  doi     = {10.1126/science.aau0685}
}

@article{Gusev2,
  author  = {Gusev, G. M. and Levin, A. D. and Levinson, E. V. and Bakarov, A. K.},
  title   = {Viscous transport and {Hall} viscosity in a two-dimensional electron system},
  journal = {Phys. Rev. B},
  volume  = {98},
  pages   = {161303(R)},
  year    = {2018},
  doi     = {10.1103/PhysRevB.98.161303}
}

@article{Scaffidi2017,
  author  = {Scaffidi, T. and Nandi, N. and Schmidt, B. and Mackenzie, A. P. and Moore, J. E.},
  title   = {Hydrodynamic electron flow and {Hall} viscosity},
  journal = {Phys. Rev. Lett.},
  volume  = {118},
  pages   = {226601},
  year    = {2017},
  doi     = {10.1103/PhysRevLett.118.226601}
}

@misc{Grigorev2025,
  author        = {Grigorev, A. A. and Afanasiev, A. N.},
  title         = {{Hall} effect in slip flow of two-dimensional electron fluid},
  year          = {2025},
  eprint        = {2505.08478},
  archivePrefix = {arXiv},
  doi           = {10.48550/arXiv.2505.08478}
}

@article{Kiselev2019,
  author  = {Kiselev, E. I. and Schmalian, J.},
  title   = {Boundary conditions of viscous electron flow},
  journal = {Phys. Rev. B},
  volume  = {99},
  pages   = {035430},
  year    = {2019},
  doi     = {10.1103/PhysRevB.99.035430}
}

@article{Afanasiev2022,
  author  = {Afanasiev, A. N. and Alekseev, P. S. and Danilenko, A. A. and Dmitriev, A. P. and Greshnov, A. A. and Semina, M. A.},
  title   = {{Hall} effect in {Poiseuille} flow of two-dimensional electron fluid},
  journal = {Phys. Rev. B},
  volume  = {106},
  pages   = {245415},
  year    = {2022},
  doi     = {10.1103/PhysRevB.106.245415}
}

@article{Raichev2022,
  author  = {Raichev, O. E.},
  title   = {Linking boundary conditions for kinetic and hydrodynamic description of fermion gas},
  journal = {Phys. Rev. B},
  volume  = {105},
  pages   = {L041301},
  year    = {2022},
  doi     = {10.1103/PhysRevB.105.L041301}
}

@article{Raichev2023,
  author  = {Raichev, O. E.},
  title   = {Magnetohydrodynamic boundary conditions for the two-dimensional fermion gas},
  journal = {Phys. Rev. B},
  volume  = {108},
  pages   = {125305},
  year    = {2023},
  doi     = {10.1103/PhysRevB.108.125305}
}

@article{Raichev2025,
  author  = {Raichev, O. E.},
  title   = {Momentum relaxation of the spin distribution function caused by electron-electron scattering in a two-dimensional Fermi gas},
  journal = {Phys. Rev. B},
  volume  = {111},
  pages   = {125308},
  year    = {2025}
}

@article{Guo2017,
  author  = {Guo, H. and Ilseven, E. and Falkovich, G. and Levitov, L. S.},
  title   = {Higher-than-ballistic conduction of viscous electron flows},
  journal = {Proc. Natl. Acad. Sci. USA},
  volume  = {114},
  pages   = {3068--3073},
  year    = {2017},
  doi     = {10.1073/pnas.1612181114}
}

@article{Afanasiev2025,
  author  = {Afanasiev, A. N. and Baryshnikov, K. A. and Korotchenkov, A. V. and Alekseev, P. S.},
  title   = {Viscoelastic resonance in two-dimensional electron flows with realistic boundary conditions at the channel edges},
  journal = {JETP Lett.},
  volume  = {122},
  pages   = {609--616},
  year    = {2025},
  doi     = {10.1134/S0021364025608693}
}

@article{Raichev2,
  author  = {Raichev, O. E. and Gusev, G. M. and Levin, A. D. and  Bakarov, A. K.},
  title   = {Manifestations of classical size effect and electronic viscosity in the magnetoresistance of narrow two-dimensional conductors: Theory and experiment},
  journal = {Phys. Rev. B},
  volume  = {101},
  pages   = {235314},
  year    = {2020},
}

@article{Alekseev2019,
  author  = {Alekseev, P. S.},
  title   = {Magnetosonic waves in a two-dimensional electron {Fermi} liquid},
  journal = {Semiconductors},
  volume  = {53},
  pages   = {1367--1374},
  year    = {2019},
  doi     = {10.1134/S1063782619100026}
}

@article{Alekseev2020,
  author  = {Alekseev, P. S. and Dmitriev, A. P.},
  title   = {Viscosity of two-dimensional electrons},
  journal = {Phys. Rev. B},
  volume  = {102},
  pages   = {241409(R)},
  year    = {2020},
  doi     = {10.1103/PhysRevB.102.241409}
}

@article{Gurzhi_J68,
  author  = {Gurzhi, R. P. and Shevchenko, S. I.},
  title   = {Hydrodynamic mechanism of electric conductivity of metals in a magnetic field},
  journal = {Sov. Phys. JETP},
  volume  = {27},
  pages   = {863},
  year    = {1968}
}

@article{Torre2015,
  author  = {Torre, Iacopo and Tomadin, Andrea and Geim, Andre K. and Polini, Marco},
  title   = {Nonlocal transport and the hydrodynamic shear viscosity in graphene},
  journal = {Phys. Rev. B},
  volume  = {92},
  pages   = {165433},
  year    = {2015},
  doi     = {10.1103/PhysRevB.92.165433}
}

\end{document}